\documentstyle[aps,epsfig,preprint]{revtex}
\def\be{\begin{eqnarray}}
\def\ee{\end{eqnarray}}
\newcommand{\bp}{\bbox{p}}
\newcommand{\bP}{\bbox{P}}

\newcommand{\bR}{\bbox{R}}

\newcommand{\un}{\underline}

\begin{document}
\tighten
\title{Phases of asymmetric nuclear matter 
with  broken space symmetries}
\author{H. M\"uther and A. Sedrakian}
\address{
Institut f\"ur Theoretische Physik, Universit\"at
T\"ubingen, D-72076 T\"ubingen, Germany
        }
\date{\today}

\maketitle
\begin{abstract}
Isoscalar Cooper pairing in isospin asymmetric nuclear matter
occurs between states populating two distinct Fermi surfaces, 
each for neutrons and protons. The transition from a BCS-like 
to the normal (unpaired) state, as the isospin asymmetry is 
increased, is intervened by superconducting phases which 
spontaneously break translational and rotational symmetries. 
One possibility is the formation of a condensate with a periodic 
crystallinelike structure where Cooper pairs carry net momentum 
(the nuclear Larkin-Ovchinnikov-Fulde-Ferrell-phase). Alternatively, 
perturbations of the  Fermi surfaces away from spherical symmetry 
allow for minima in the condensate free energy which correspond to 
a states with quadrupole deformations of Fermi surfaces and zero 
momentum of the Cooper pairs. In a combined treatment of these phases 
we show that, although the Cooper pairing with finite momentum might 
arise as a local minimum, the lowest energy state features are
deformed Fermi surfaces and Cooper pairs with vanishing total momentum.
\end{abstract}

\section{Introduction}

The pairing properties of nuclear systems - finite nuclei and
the bulk nuclear matter - play an important role 
in the physical manifestations of these systems. In most cases 
of interest the pairing occurs at finite isospin asymmetry. 
For example, charge neutral bulk nuclear 
matter with an admixture of leptons in $\beta$ equilibrium 
implies proton abundancies at about $40\%$
in supernova matter and $10\%$ in neutron star matter. 
The dominant partial-wave channel for the pairing depends on the 
density, temperature, and the isospin asymmetry in general. 
For large enough asymmetries  the matter  is paired in the  
isospin triplet channels - the $^1S_0$ channel at low 
densities
\cite{CLARK1,NISKANEN,AMU85,CHEN86,AINS89,BCLL90,WAMBACH,KHODEL96,SCHU,ELGA98,KUK}  
and $^3P_2$-$^3F_2$ or $^1D_2$ channels at high densities\cite{RUNDERMAN,TAKATSUKA1,TAKATSUKA2,RICHARDSON,AMU285,ELGA96,BEEHS98,KHODEL98,KHODEL2001}.
For weakly asymmetric systems the 
isospin singlet attractive
$^3S_1$-$^3D_1$\cite{{SD_PAIRING3,NUCL_BCS1,NUCL_BCS2,NUCL_BCS3,MORTEN,NUCL_BCS4}}
and $^3D_2$ channels\cite{SED95,SED96}
dominate the pairing interaction at low and high densities, respectively. 
In bulk nuclear matter of compact stars the
asymmetries are most likely too large to support isospin singlet
pairing. However the dilute nuclear matter in supernovas and the
low-density tails  of the exotic nuclei
can support $^3S_1$-$^3D_1$-channel pairing, since in general 
the isospin asymmetry is less effective in destroying the pair
correlations in dilute matter\cite{NUCL_BCS3}.

The mechanism of the suppression of the BCS pairing of the kind found  
in asymmetric nuclear matter  was initially 
studied in metallic superconductors in a spin-polarizing magnetic
field\cite{{AG,CLOGSTON,CHANDRA,GR}}.
Since a homogenous magnetic field
is screened beyond the London penetration depth in the bulk of a 
superconductor, in real materials the electron
spin polarization is induced by low-density
paramagnetic impurities, which interact with an electron by spin
nonconserving forces.
The effect of collision induced polarization
is commonly modeled by an average polarizing field which implies an
asymmetry in the populations of spin-up and -down quasiparticles.
The resulting electron  Pauli paramagnetism suppresses  the $S$-wave 
pairing  when the Fermi surfaces of the electrons 
with spin-up and -down are apart on a scale of the order of 
the energy gap in the quasiparticle spectrum. A close analogy to the
ordinary superconductors can be seen in the effect of the large 
$B$ fields in the neutron stars on the $S$-wave pairing of neutrons.
Our general discussion below also applies to the pairing in 
spin-polarized neutron matter in the $^1S_0$ channel, although the numerical
computations are carried out for the isoscalar pairing in asymmetric 
matter.

The understanding of the 
isospin singlet pairing as a function of the isospin  asymmetry 
requires an extension of the standard BCS theory to
describe phases with spontaneously broken space symmetries.  
Larkin and Ovchinnikov\cite{LO} 
and Fulde-Ferrell\cite{FF} (LOFF) argued, in the context
of metallic superconductors, that a transition from the BCS paired 
to the normal state occurs via a phase where the Cooper pairs 
carry a finite total momentum.  
The isospin polarization in nuclear matter 
with pairing in the $^3S_1$-$^3D_1$ channel leads to a nuclear LOFF phase 
for a range of isospin asymmetries\cite{NUCL_LOFF,ISAEV}. Unlike the metallic
superconductors where the ratio of the pairing gap to the average
chemical potential is of the order 10$^{-3}$,   
the nuclear LOFF phase emerges in the strong coupling
regime, where this ratio is of the order 0.1.  
The spatially periodic structure 
of the nuclear LOFF phase corresponding to oscillations of the 
condensate wave function are on the scales of the order of several fm.
Note that, independent of the detailed
structure of the lattice, the LOFF phase breaks {\it both} the
translational and rotational invariance of the original BCS state. 
The formation of the LOFF phase is a robust feature of the 
fermionic systems with broken time-reversal symmetry and has been 
studied in detail in the flavor asymmetric high-density QCD 
(crystalline color superconductivity)
\cite{Alford:2001dt,Rajagopal:2001yd,Rajagopal:2000wf,Alford:2000ze,Leibovich:2001xr,Bowers:2001ip,Casalbuoni:2002pa,rajagopal:2002}.

The evolution of the condensate from the BCS to paired states with
broken space symmetries, as the isospin asymmetry is increased, can take 
a different path if we give up the assumption that the intrinsically
homogeneous  matter features spherical Fermi surfaces\cite{MS}.
As shown in ref. \cite{MS}, for quadrupole deformed 
Fermi surfaces (DFS) the BCS condensate evolves to a state with
spontaneously broken rotational symmetry [in group theory terms
the $O(3)$ symmetry breaks down to $O(2)$]. Such a state is still 
axially symmetric and, unlike the LOFF phase, preserves the
translational symmetry. We shall refer to 
the superconducting state with deformed Fermi surfaces  as the
DFS phase with the understanding that the expansion describing the
deformation of Fermi surfaces is truncated at 
quadrupole order (the higher order terms in the
multiple expansion have not been studied so far).

Why do the phases with broken space symmetries dominate the BCS
state? The BCS quasiparticle spectrum (for homogeneous systems)
is isotropic; when 
the polarizing field drives apart the Fermi surfaces of fermions
with spin/isospin up and down the phase space overlap is lost, 
hence the pairing is suppressed.
The finite momentum of the Cooper pairs of the LOFF phase 
implies an anisotropic
quasiparticle spectrum; the magnitude of the anisotropy is
controlled by the value of the net momentum of the pairs which is 
treated as a parameter. This new degree of freedom, which can be
viewed as an additional variational parameter for the minimization of
the ground state of the system,  (partially) overcomes  the
phase-space loss caused by the departure of the Fermi surfaces from
the perfectly overlapping configuration. In
effect the gain in the pairing energy through increase of the pairing 
field dominates  the increase of the
kinetic energy. To motivate the  DFS phase, note that
formally the LOFF spectrum can be viewed as a dipole ($\propto P_1(x)$)
perturbation of the spherically symmetrical
BCS spectrum,  where $P_l(x)$ are the Legendre polynomials, and $x$ is the
cosine of the angle between the particle momentum and the total
momentum of the Cooper pair. The DFS phase  assumes an expansion 
of the quasiparticle spectrum up to the next-to-leading order,
quadrupole ($\propto P_2(x)$)   deformation; it also assumed that 
the total momentum  of the Cooper pairs is zero, i.e., there are
quadrupole deformations only.
Although the perturbations from the spherical symmetry cost kinetic 
energy, they are compensated by the gain in the potential energy. For 
a range of asymmetries the DFS phase becomes the ground state of the 
paired system\cite{MS}. 

The purpose of the present paper is a combined  treatment of the 
LOFF and the DFS phases in bulk isospin asymmetric nuclear matter 
within the same model as that in Refs. \cite{NUCL_LOFF,MS}. 
The model assumes a pairing force given by a
phase-shift equivalent (i.e., realistic) interaction and
nonrenormalized quasiparticle spectrum. 
The effect of the mean field in the
Brueckner-Hartree-Fock approximation is 
to reduce the range of isospin asymmetries where 
the pairing exists\cite{NUCL_BCS2}. We shall follow the strategy of
disentangling the effects of the mean field from the pairing since
already at the level of approximations of Refs. \cite{NUCL_LOFF,MS} the
LOFF and DFS phases show a complicated behavior. The renormalization 
of the quasiparticle spectrum  for the pairing interaction
is a problem of its own which has not been solved to date.
Issues such as the consistency between the 
vertex and propagator renormalization and fulfillment of the spectral sum rules,
etc. need further attention.

In Section 2 we derive the BCS equations, which include the effects
of the finite momentum of the Cooper pairs and Fermi surface
deformations, within the finite-temperature real-time Green's functions formalism.
The numerical solutions of these equations are presented in Section 3,
where we discuss the phase diagram of the combined LOFF and DFS phases
and their properties at finite temperatures. 
Section 4 contains a summary of the results and an outlook.

\section{Formalism}
The starting point of the model is the Hamiltonian of fermions
interacting via two-body forces:
\be \label{H}
H&=&\frac{1}{2m}\sum_{\alpha}\int d^3r \vec \nabla\psi^{\dagger}(\vec r,t,\alpha)
\vec \nabla\psi(\vec r,t,\alpha)  \nonumber \\
&+&\frac{1}{2}\sum_{\alpha\beta}
\int d^3r d^3r' \psi^{\dagger}(\vec r,t,\alpha)
\psi^{\dagger}(\vec r',t,\beta)
V(\vec r,t,\alpha;\vec r',t',\beta)
\psi(\vec r,t,\beta) \psi(\vec r,t,\alpha),
\ee
where $\psi(\vec r,t,\alpha)$ are second 
quantized operators in  the Heisenberg representation, 
$V_{\alpha,\beta}(\vec r-\vec r')\delta(t-t')$ is 
the space and time local bare interaction, and $\alpha$, $\beta$ stand for
discrete quantum numbers (spin, isospin, etc.).
We choose to formulate the finite-temperature pairing theory 
in terms of nonequilibrium real-time Green's functions.
The single-particle propagator in this formalism
is combined in a $2\times 2$ matrix, 
\be\label{2}
\underline{\hat G}(x_1,x_2) =\left(\begin{array}{cc} \hat G^{c}(x_1,x_2)& 
\hat G^{<}(x_1,x_2)\\
\hat G^{>}(x_1,x_2)& \hat {G}^{a}(x_1,x_2)\end{array}\right)
=\left(\begin{array}{cc} -i\langle T\,\hat  \psi(x_1)\, \hat\psi^{\dagger}(x_2)\rangle
  & i\langle \, \hat \psi^{\dagger}(x_2)\, \hat
\psi(x_1)\rangle\\ -i\langle \, \hat\psi
  (x_1)\,\hat \psi^{\dagger}(x_2)\rangle & -i\langle \tilde T\, \hat\psi(x_1)\,
 \hat \psi^{\dagger}(x_2)\rangle\end{array}\right),
\ee
where $x\equiv(\vec r, t, \alpha)$, and the averaging is over an arbitrary 
nonequilibrium state of the system; $T$ and $\tilde T$ are
chronological and antichronological time ordering operators for the 
Nambu spinors
\be\label{NAMBU}
\hat\psi^{\dagger} = \left[\psi^{\dagger}(x)\psi(x)\right]
\quad\hat\psi = \left[\begin{array}{c} \psi_(x)\\
\psi^{\dagger}(x)\end{array}\right] .
\ee   
The Nambu spinors span the particle-hole space and satisfy the 
fermionic equal-time anticommutation relations
\be 
\{\hat\psi_{\alpha}(\vec x,t),\hat\psi_{\beta}^{\dagger}(\vec x',t)\}
=\delta_{\alpha\beta}\delta^3(\vec x-\vec x'),\quad
\{\hat\psi_{\alpha}(\vec x,t),\hat\psi_{\beta}(\vec x',t)\}=0,\quad 
\{\hat\psi_{\alpha}(\vec x,t)^{\dagger},\hat\psi_{\beta}^{\dagger}(\vec x',t)\}
=0.\nonumber
\ee               
Starting from the equations of motion for the Heisenberg 
operators and the Hamiltonian (\ref{H})  the 
Martin-Schwinger hierarchy can be constructed 
for a matrix Green's function involving an arbitrary number of 
field operators.
For our purposes in is sufficient to truncate the hierarchy
at the $n=1$ level. The equation of motion of a one-particle 
Greens function is then given by the time-dependent Dyson equation
\be\label{DYSON01}
\un{\hat G}_{(0)}^{-1}(x_1)\otimes\un{\hat G}(x_1,x_2) = 
\sigma_z\delta(x_1-x_2)
           - \un{\hat V}(x_1,x_6;x_5,x_4)\otimes
\un{\hat{G}}(x_5,x_4,x_2,x_6^+),
\ee  
where the subscript $(0)$ refers to the free propagator, $\otimes$
stands for summation/integration over the repeated discrete/continuous
variables, $\sigma_i$ are the components of the vector of the Pauli
matrices, and the  superscript $+$ means an infinitesimal increment in the time
argument. The hierarchy at a given level can be decoupled (formally)
by introducing the self-energy matrix:
\be\label{DYSON2}
\un{\hat G}_{(0)}^{-1}(x_1)\otimes\un{\hat G}(x_1,x_2) =\sigma_z\delta(x_1-x_2)
           +  \un{\hat\Sigma}(x_1,x_3)\otimes
             \un{\hat{G}}(x_3,x_2),
\ee
which is defined as 
\be\label{SIGMA1}
\un{\hat\Sigma}(x_1,x_2) = - 
\un{\hat V}(x_1,x_6;x_5,x_4)\otimes\un{\hat{G}}(x_5,x_4,x_7,x_6^+)
\un{\hat G}^{-1}(x_7,x_2).
\ee
The first of these equations defines the quasiparticle 
spectrum of the superconducting state, the second,
the self-energies, e.g., the gap function. To 
close the set of exact equations we need to specify the 
approximation to the two-particle Green's function. The BCS theory
follows in the Hartree approximation, where the two-particle Green's
function is approximated by a product of single-particle Green's 
functions. Note that the anomalous contributions of the type 
$\langle \psi(x_1)\psi(x_2)\rangle $
and $\langle \psi^{\dagger}(x_1)\psi^{\dagger}(x_2)\rangle $
are automatically built in in the Hartree approximation.
This can be verified from the 
explicit form of the propagator in Eq. (\ref{2}), when the 
spinors (\ref{NAMBU}) are substituted. Each element of the $2\times 2$
matrix  (\ref{2}) is in turn a
$2\times 2$ matrix in the particle-hole 
space:
\be\label{1}
{\hat G}_{12}^j =\left(\begin{array}{cc} {G}^j(x_1,x_2)& 
{F}^j(x_1,x_2)\\ 
{F}^{\dagger\, j}(x_1,x_2)& \overline{G}^j (x_1,x_2)
\end{array}\right),
\ee
where $j\equiv c,a,>,<$. 
The matrix form of the self-energies in the particle-hole space
is analogous to that of the propagators:
\be 
\underline{\hat\Sigma}(x_1,x_2) =
\left(\begin{array}{cc} \hat\Sigma^{c}(x_1,x_2)& \hat\Sigma^{<}(x_1,x_2)\\
\hat\Sigma^{>}(x_1,x_2)& \hat\Sigma^{a}(x_1,x_2)\end{array}\right),
\quad 
\hat \Sigma^j (x_1,x_2)\equiv \left( \begin{array}{cc}
        \Sigma^j (x_1,x_2) &\Delta^j(x_1,x_2) \\
        \Delta^{\dagger\,j} (x_1,x_2)
         & \overline\Sigma^j(x_1,x_2)
      \end{array} \right),
\ee
where the off-diagonal elements of the $\Sigma^j (x_1,x_2)$ matrix
correspond to the pairing amplitude.
Note that the inverse free-particle propagator is diagonal in 
the particle-hole, $j$ and spin/isospin spaces:
\be 
{G}^{-1}_{(0)\alpha\beta} = 
\left(i\partial_t + \nabla^2/2m_{\alpha}-\mu_{\alpha}\right)
\delta_{\alpha\beta}.
\ee
The Dyson
Eq. (\ref{DYSON2}) is a $16\times16$ matrix equation in 
general if we include the spin and isospin. The size of the matrix 
to be diagonalized can be reduced in several successive steps. Here
we note that in the
time-independent stationary limit the number of the propagators
can be reduced by applying the unitary transformation 
$U = (1+i\sigma_y)/\sqrt{2}$ to Eqs. (\ref{DYSON2}) and 
(\ref{SIGMA1}),  in the $j$ space, and 
introducing the retarded/advanced propagators by the relations 
\be\label{RETARDED}
G^{R,A}(x_1,x_2) = G^{c}(x_1,x_2) - G^{<,>}(x_1,x_2)
= G^{>,<}(x_1,x_2) - G^{a}(x_1,x_2).
\ee 
Further reductions will be carried out in the next section.
The Dyson equation, e.g., for the retarded propagator is invariant
under the rotation in the $j$ space affected by the transformation 
above, i.e.,
\be\label{DYSON3}
{\hat G}^R(x_1,x_2) ={\hat G}_{(0)}^R(x_1,x_2)+
{\hat G}_{(0)}^R(x_1,x_2)\otimes{\hat\Sigma}^R(x_1,x_3)
\otimes{\hat{G}}^R(x_3,x_2);
\ee
the retarded self-energies are related to the matrix components in
the $j$ space by relations analogous to Eq. (\ref{RETARDED}). 
The integral Eq.  (\ref{DYSON3}) for the retarded propagators can
be solved without specifying the approximation to (and hence the form of)
the self-energies, in the quasiclassical approximation. 

\subsection{Quasiparticle spectrum}
When the  characteristic length scales of the spatial variations of 
the macroscopic condensate are much larger than the inverse of the 
momenta involved in the problem $\sim p_F$, where
$p_F$ is the Fermi momentum, the two-point correlation functions can 
be approximated by their quasiclassical counterparts. Going over to the
center of mass $X = (x_1+x_2)/2$ and relative $\xi = x_1-x_2$ coordinates in
the two-point functions and carrying a Fourier transform with respect
to the relative coordinates we arrive at the mixed 
representation for these functions, e.g.,
\be\label{MIX}
\hat G^{<}(p, X)=\int\!e^{ip\xi}~\hat G^{<}\left(X+\frac{\xi}{2},X-\frac{\xi}{2}\right)d^4\xi ,
\ee 
where $p\equiv(\omega, \bp)$ are the relative frequency and momentum, 
and $X \equiv (\bR, T)$.  Since the
variations of the propagators and self-energies are slow on the scales
of the order of $\bR$, keeping the leading order terms in the
gradient expansion is accurate to order $\sim {\cal O}(p_FR$).
The quasiclassical counterpart of the Dyson Eq.  (\ref{DYSON3})
is
\be\label{QC}
\sum_{\gamma}\left( \begin{array}{cc}
\omega - \epsilon_{\alpha\gamma}^+
       &-\Delta_{\alpha\gamma}^R(\omega, \bp) \\
       - \Delta_{\alpha\gamma}^{\dagger R}(\omega, \bp)
& \omega + \epsilon_{\alpha\gamma}^-
      \end{array} \right) \left( \begin{array}{cc}
        G_{\gamma\beta}^R(\omega, \bp)& F_{\gamma\beta}^R(\omega, \bp) \\
        F_{\gamma\beta}^{\dagger, R}(\omega, \bp)
        & \overline G_{\gamma\beta}^{R}(\omega, \bp)
      \end{array} \right)= \delta_{\alpha\beta} \hat {\bf 1},
\ee
where
\be\label{SP}
\epsilon_{\alpha\beta}^{\pm}=
 [\left(\bP/2\pm\bp\right)^2/2m_{\alpha}-\mu_{\alpha}
]\delta_{\alpha\beta}\nonumber
&\pm& \Sigma^{R,(+)}_{\alpha\beta}(\omega,\bP/2\pm\bp ) 
-\Sigma^{R,(-)}_{\alpha\beta}(\omega, \bP/2\pm\bp),
\ee
and $\hat {\bf 1}$ is a unit matrix in the particle-hole space.
Here we have defined the symmetric and antisymmetric combinations 
of the particle/hole retarded self-energies
\be 
\Sigma_{\alpha\beta}^{R,(+)}(\omega, \bp) &\equiv&\frac{1}{2}
\left[\Sigma^R_{\alpha\beta}(\omega,
\bp)+\overline\Sigma^R_{\alpha\beta}(\omega, \bp)\right],\\
\Sigma^{R,(-)}_{\alpha\beta}(\omega, \bp) &\equiv&\frac{1}{2}
\left[\Sigma^R_{\alpha\beta}(\omega,
\bp)-\overline\Sigma^R_{\alpha\beta}(\omega, \bp)\right].
\ee
The dependence on the center-of-mass time is dropped in the above 
expressions in the stationary limit. The retarded self-energies
now can be expanded in Legendre polynomials:
\be\label{EXP1}
\Sigma^{R,(+/-)}_{\alpha\beta}(\omega,\bp ) &=& 
\sum_{l = 0}^{\infty}\Sigma^{R,(+/-)}_{\alpha\beta, 2l}(\omega,p)
P_l({\rm Cos}~\theta),
\ee
where $\theta$ is the angle between the vectors $\bp$ and $\bP$. 
In practice the expansion will be truncated at order $l=2$; 
the $l=0$ term renormalizes the chemical potential, as discussed below.
For translationally invariant interactions the odd terms 
do not contribute in the expansion (\ref{EXP1}); the interactions 
are assumed to be timelocal. 
For translationally noninvariant interactions the effect of the 
$l=1$ term would be
a renormalization of the dipole deformation in the spectrum due to the
kinetic energy terms. It can be included in the renormalization of
the particle mass entering the anisotropic contribution to the kinetic
energy, although it would not contribute to the net isotropic increase
of the kinetic energy due to the finite momentum.
Since the total momentum is treated as a variational parameter
in the problem, the effect of the  self-energy
renormalization at $l=1$ can be accounted for by a simple rescaling of
this parameter; therefore we anticipate 
that the $l=1$ terms would not change the results below.

For spin and isospin conserving forces the normal
Green's functions and self-energies are diagonal in the spin and
isospin spaces. It is sufficient to consider the anomalous propagators,
e.g., in the isospin space, since the resulting spin structure,
for $S$-wave interactions, is uniquely determined for each isospin
combination.  If we restrict ourselves to the neutron-proton pairing in
the $^3S_1$-$^3D_1$ channel, which is justified when
$\Delta_{nn},\Delta_{pp} \ll \Delta_{np}$ 
(subscripts $n$ and $p$ refer to protons and neutrons)
then $\Delta_{\alpha\beta}=\sigma_x\Delta$.
The spectrum, in this case, takes the 
form
\be\label{SPECTRUM2}
\omega_{\pm} = E_A \pm \sqrt{E_S^2+\vert \Delta\vert^2},
\ee
where the symmetric and asymmetric
parts of the spectrum (which are even and odd
with respect to the time-reversal symmetry) are defined as
\be 
E_S = \frac{1}{2}(\epsilon_++\epsilon_-), \quad 
E_A = \frac{1}{2}(\epsilon_+-\epsilon_-).
\ee
The eigenvalues (\ref{SPECTRUM2}) are valid for an arbitrary approximation 
to the self-energies. Further,  we shall approximate the symmetric and
antisymmetric parts of the spectra using  (i) the  quasiparticle
and  (ii) the momentum-independent self-energy shift
approximations. The first approximation implies 
\be \label{QPA}
\Sigma_{n/p}^{R,(-/+)}(\bp,\omega) \simeq
{\rm Re}\Sigma_{n/p}^{R,(-/+)}(\bp,\epsilon_p) \simeq 
 {\rm Re}\Sigma_{n/p}^{R,(-/+)}(\bp,\epsilon_p)\Big\vert_{p_F} 
+\frac{\partial }{\partial \bp}{\rm Re}\Sigma_{n/p}^{R,(-/+)}(\bp,\epsilon_p)
\Big\vert_{p_F} (p-p_F);
\ee
the second approximation keeps the leading order (constant) term in
the expansion (\ref{QPA}).
Within these approximations we obtain
\be
\label{ES}
E_S &=&\frac{P^2}{8m}+\frac{p^2}{2m}-\mu -(\mu\epsilon +
\delta\epsilon\delta\mu) \cos^2\theta ,
\\ 
\label{EA}
E_A &=&\frac{Pp}{2m} \cos\theta -\delta\mu 
+\left(\mu\delta\epsilon+\epsilon\delta\mu\right)\cos^2\theta ,
\ee
where we defined 
the average chemical potential and the conformal deformation
\be\label{MU1}
\mu &=&\frac{1}{2}\left[\mu_{n}+\mu_p\right]
-\frac{1}{2}
\left[\Omega^-_n(p_F)+\Omega^-_p(p_F)\right],\\
\label{MU2}
\epsilon &=& \frac{3}{4\mu}{\rm Re}\left[\Sigma^{R(-)}_{n,2}(p_F)+\Sigma^{R(-)}_{p,2}(p_F)
\right],
\ee
as the $l=0$ and $l=2$ contributions from the isospin
symmetric part of the quasiparticle spectra; similarly the relative 
chemical potential and the relative deformation are defined as 
the $l=0$ and $l=2$ contributions from the isospin
antisymmetric part of the quasiparticle spectra:
\be 
\label{MU3}
\delta\mu &=& \frac{1}{2}\left[\mu_{n}-\mu_p\right]
-\frac{1}{2}\left[\Omega^+_n(p_F)-\Omega^+_p(p_F)\right],\\
\label{MU4}
\delta\epsilon &=& \frac{3}{4\mu}{\rm Re}\left[\Sigma^{R(+)}_{n,2}(p_F)
-\Sigma^{R(+)}_{p,2}(p_F)
\right],
\ee
where we used the abbreviation
\be
\Omega^{-/+}_{n/p}(p_F)={\rm Re}\left[
\Sigma^{R(-/+)}_{n/p,0}(p_F)-\frac{1}{2}\Sigma^{R(-/+)}_{n/p,2}(p_F)
\right].
\ee

The self-energies on the right-hand side of Eqs. (\ref{MU1})-(\ref{MU4})
are functionals of the pairing field, i.e., they  are nonzero even in the
case where all the interactions apart from the pairing interaction
are switched off. They depend on the quasiparticle momentum via the 
momentum dependence of the gap function; if the latter is approximated
by its value at the Fermi momentum the self-energies become 
constants. The values of the average and relative chemical 
potentials  $\mu$ and $\delta\mu$ can be   
adjusted to reproduce the particle density at a given
temperature and isospin asymmetry. 
The conformal and relative deformations 
$\epsilon$ and $\delta\epsilon$ along with the finite
momentum of the pairs $P$ are treated as variational 
parameters to be determined from the ground state of the
superconducting system. 

The solution of the Dyson Eq. (\ref{QC}) for the propagators can 
be written in terms of the eigenvalues as
\be
G_{n/p}^R(\omega, \bp) &=& \frac{u_p^2}{\omega-\omega_{+/-}+i\eta}
  +  \frac{v_p^2}{\omega-\omega_{-/+}+i\eta}, \\
F^R(\omega, \bp)&=& u_p v_p \left( \frac{1}{\omega-\omega_{+}+i\eta}
      -\frac{1}{\omega-\omega_{-}+i\eta}\right),
\ee
where the Bogolyubov amplitudes are
\be
u_p^2 = \frac{1}{2} + \frac{E_S}
          {2\sqrt{E_S^2+\vert \Delta\vert^2}} , \quad
v_p^2 = \frac{1}{2} - \frac{E_S}
          {2\sqrt{E_S^2+\vert \Delta\vert^2}} .
\ee
The remainder Green's functions can be reconstructed using the 
spectral representation of the retarded Green's functions 
\be \label{SPforG}
G_{n/p}^R(\omega, \bp) &=& \int_{-\infty}^{\infty}\! d\omega'
\frac{A_{n/p}(\omega', \bp)}{\omega-\omega'+i\eta},\\
\label{SPforF}
F^R(\omega, \bp)&=& \int_{-\infty}^{\infty}\! d\omega'
\frac{B(\omega', \bp)}{\omega-\omega'+i\eta},
\ee
where $ A_{n/p}(\omega, \bp)$ and $B(\omega, \bp)$  are the spectral 
functions. The Kadanoff-Baym ansatz\cite{KB} provides the link between the 
the retarded components and the remainder Green's 
functions in the $j$ space.
We have 
\be \label{KBforG}
G_{n/p}^{>,<}(\omega, \bp) &=& A_{n/p}(\omega, \bp) f^{>,<}(\omega),
\\
\label{KBforF}
F^{>,<}(\omega, \bp) &=& B(\omega, \bp) f^{>,<}(\omega),
\ee
where, in equilibrium, the Wigner distribution functions of the
Kadanoff-Baym ansatz\cite{KB}
reduce to the Fermi distribution function 
$f^<(\omega)=\left[{\rm exp}(\beta\omega)+1\right]^{-1}$ and
$f^>(\omega)=\left[1-f^<(\omega)\right]$;
here  $\beta$ is the inverse temperature. Note that the relations 
above are exact in the equilibrium limit.
The causal and acausal
Green's functions are then obtained from the relations (\ref{RETARDED}).

\subsection{The gap equation}
The BCS gap equation follows from the Hartree approximation for the
two-particle Green's function  in Eq. (\ref{DYSON01}):
\be
\underline{\hat G}(x_1,x_2;x_3,x_4) = \underline{\hat G}(x_1,x_2)
\otimes \underline{\hat G}(x_3,x_4),
\ee
which implies, according to Eq. (\ref{SIGMA1}),
\be 
\underline{\hat \Sigma}(x_1,x_2) = - \underline{\hat V}(x_1,x_3;x_2,x_4)
\otimes\underline{\hat G}(x_4,x_3^+).
\ee
Upon applying the quasiclassical approximation to the above equation
in the case of time local interaction, we obtain  the retarded 
(off diagonal in the particle-hole space) self-energy
\be\label{GAP2}
\Delta^R (\bp,\bP) = \int \frac{d^3p'd\omega'}{(2\pi)^4}
V(\bp,\bp') F^< (\omega',\bp',\bP),
\ee
or using the relation 
$F^< (\omega,\bp) = -2 {\rm Im} F (\omega,\bp)f(\omega)$, 
which follows from Eqs. (\ref{SPforF}) and (\ref{KBforF}),
\be\label{GAP3}
\Delta^R (\bp,\bP) = -\int \frac{d^3p'd\omega'}{(2\pi)^4}
V(\bp,\bp') {\rm Im}F^R(\omega',\bp',\bP)f^<(\omega') .
\ee
Further progress requires partial-wave
decomposition of the interaction, which can be performed after
angle averaging  the remainder functions on the
right-hand side of Eq. (\ref{GAP3}).
The result of this procedure is
\be\label{GAP4}
\Delta_l^R(p,P)=-\sum_{l'} \int \frac{dp p^2}{(2\pi)^2} V_{ll'}(p,p')
\frac{\Delta_{l'}^R(p',P)}{2\sqrt{\epsilon_S^2+ \Delta(p',P)^2}}
\langle \left[f^<(\omega_+)-f^<(\omega_-)\right]\rangle ,
\ee
where  $\langle \dots \rangle $ denotes the average over the angle between
the relative and total momenta and $\Delta(p,P)^2 \equiv \Delta_0(p,P)^2
+ \Delta_2(p,P)^2$ is the angle averaged gap. Here the pairing
interaction is approximated
by the bare neutron-proton interaction $V(\bp,\bp') $
in the $^3S_1$-$^3D_1$ channel.
As discussed above the average and relative chemical potentials 
can be fixed by adjusting the pairing field to reproduce the
matter density $\rho = \rho_n+\rho_p$ and the isospin asymmetry 
$\alpha = (\rho_n-\rho_p)/\rho$. The 
total momentum,  the relative and conformal deformations, 
are treated as variational parameters to be determined from the 
ground state energy of the system.
The corresponding expression for partial densities of protons and 
neutrons is provided by the relation
\be 
\rho_{n/p}=\int \frac{d^4 p}{(2\pi)^4} G_{n/p}^<(\omega,\bp),
\ee
or, using Eqs. (\ref{SPforG}) and (\ref{KBforG}),
\be\label{DENSITY}
\rho_{n/p}= -2\sum_{\sigma}\int \frac{d^4 p}{(2\pi)^4}{\rm Im}
G_{n/p}(\omega,\bp) f(\omega) = \sum_{\sigma}\int \frac{d^3p}{(2\pi)^3}
\left\{u_p^2 f^<(\omega_{\pm})+v_p^2 f^<(\omega_{\mp})\right\},
\ee
where $\sigma$ stands for quasiparticle spin and the second relation 
follows in the quasiparticle approximation.

\subsection{Thermodynamics}
The phase diagram of the paired state at fixed finite temperature and 
density can be obtained from the free energy in the mean field
approximation. The  free energy is given by the thermodynamic relation
\be\label{GR}
{\cal F}\vert_{\rho,\beta} =  {\cal U} - \beta^{-1}S.
\ee
where ${\cal U}$ is the internal energy and $S$ is the entropy. 
A thermodynamically stable paired state minimizes
the difference of the free energies of the superconducting 
and normal states, $\delta {\cal F} \equiv {\cal F}_S-{\cal F}_N$. 
The mean field entropy of the superfluid state is
\be\label{ENTROPY}
S_S = - 2 k_B \sum
\Bigl\{f^<(\omega_{+})\,{\rm ln}\, f^<(\omega_{+})+
         f^>(\omega_{+})\,{\rm ln}\, f^>(\omega_{+})
+f^<(\omega_{-})\,{\rm ln}\, f^<(\omega_{-})
+f^>(\omega_{-})\,{\rm ln}\, f^>(\omega_{-})
\Bigr\},
\ee
where  $k_B$ is the Boltzmann constant and the sum is over the momentum
states in the quasiparticle approximation.
The corresponding expression for the entropy of the 
normal state $S_N$ is obtained by taking the 
limit $\Delta \to 0$ in Eq. (\ref{ENTROPY}). 
The internal energy of the superconducting state in the mean field
approximation is
\be\label{U}
{\cal U}_S= 2\int \frac{d^3p}{(2\pi)^3}\Biggl\{
\left[\epsilon^+n_n(p)+\epsilon^- n_p(p) \right]
+\sum_{ll'}\frac{d^3p'}{(2\pi)^3}
\, V_{ll'}(p,p')\, \nu_l(p)\nu_{l'}(p')\Biggl\},
\ee
where the normal and superconducting occupation probabilities are 
defined as 
\be
n_{n/p}(p)\equiv u_p^2 f^<(\omega_{\pm})+v_p^2 f^<(\omega_{\mp}) ,
\quad
\nu(p) \equiv u_pv_p \left[f^<(\omega_+)
-f^<(\omega_-)\right].
\ee
The first term in Eq. (\ref{U}) includes the kinetic energy of 
quasiparticles which is a functional of the pairing gap. In the normal
state it reduces to the kinetic energy of noninteracting quasiparticles.
The second term describes the BCS mean-field interaction among the 
particles in the condensate and vanishes in the normal state.

\section{Results}

The nuclear LOFF and DFS phases were studied numerically using the
Paris nucleon-nucleon interaction. Our results did not depend on 
this  choice of the interaction, since there are no
significant deviations in the  scattering phase shifts in the
$^3S_1-^3D_1$ channel among various realistic potentials 
and these reproduce the  properties of the deuteron at the same level
of accuracy. The net density of the matter was fixed at the empirical 
saturation density of the nuclear matter $\rho_s = 0.17$ fm$^{-3}$. 
At this density the typical energies of the nucleons are below the
threshold laboratory energy $E_{LAB}  = 220 $ MeV at which  
the interaction in the $^3S_1-^3D_1$ channel becomes repulsive.
Lower densities could provide a more realistic (from the physical point of
view) setting, however, they have the disadvantage that the effects of the
Bose-Einstein condensation of deuterons may start to play a role\cite{NUCL_BCS3}. 
The computations were carried out for several fixed values of the
density asymmetry $\alpha$ that are above the critical values at which the 
nuclear LOFF and DFS phases set in (see Refs. \cite{NUCL_LOFF,MS}). The
qualitative behavior of quantities of interest is generic 
for different values of $\alpha$, therefore, we shall show the results
for a fixed value  $\alpha = 0.35$ at which the LOFF and DFS phases 
(separately)  dominate the ordinary BCS state.
\begin{figure}[htb] 
\begin{center}
\includegraphics[angle=-90,width=\linewidth]{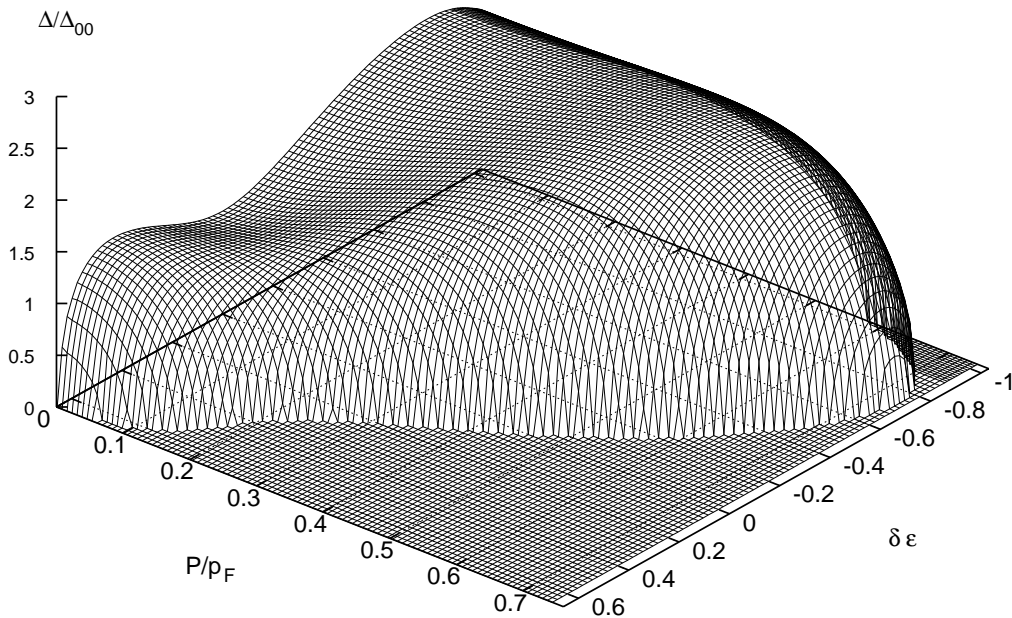}
\end{center}
\caption{The pairing gap as a function of a Cooper pair momentum $P$
in units of Fermi momentum and the relative deformation parameter 
$\delta\epsilon$. The density asymmetry is fixed at the value 
$\alpha =0.35$. The temperature and the density are 
$T = 3$ MeV and $\rho_s = 0.17$ fm$^{-3}$, respectively. The gap is
normalized to its value $\Delta_{00}$ in the asymmetric BCS state.
}
\label{MSfig:fig1}
\end{figure} 
\noindent
The results of the combined treatment of the LOFF and DFS phases
will be presented in two steps: in the next section we fix temperature  and
vary simultaneously the net momentum of the pairs $P$ (LOFF phase) and 
the relative deformation $\delta\epsilon$ (DFS phase). In the
subsequent section we study the temperature dependence of these 
phases at either constant $P$ or $\delta\epsilon$.

\subsection{LOFF versus DFS}
Figure 1 shows the pairing gap as a function of the total momentum of
Cooper pairs $P$ and the relative deformation $\delta\epsilon$ 
for vanishing conformal deformation ($\epsilon = 0$). 
The temperature is fixed at the value $T = 3$ MeV. 
The ratios of the temperature over the Fermi energy and the pairing
gap in the symmetric state are 0.1 and 0.4 respectively,
i.e., they correspond to the low-temperature regime. 
The gap is normalized to its value
$\Delta_{00}\equiv \Delta(P=0,\delta\epsilon=0)$  in the
rotationally/translationally invariant state at $\alpha = 0.35$.
Although $\alpha \equiv (\rho_n-\rho_p)/(\rho_n+\rho_p)$ 
changes in the interval $[-1;1]$ in general, 
the symmetry of the equations with respect to the indices 
labeling the species reduces the range of $\alpha$  to $[0;1]$. 
The relative deformation obviously 
is not bounded and can assume both positive and negative
values. Positive values of $\delta\epsilon$  imply an
oblate deformation for the Fermi surface of neutrons and a prolate
deformation for the Fermi surface of protons.
The behavior of the pairing gap can be understood by examining 
the effect of the symmetric and antisymmetric parts of the
quasiparticle spectrum on the gap Eq. (\ref{GAP4}). The
antisymmetric part $E_A$ appears only in the Fermi distribution
functions. If  $E_A= 0$ one recovers the BCS limit, where
the Fermi distribution functions become identical, i.e., the 
Fermi surfaces perfectly coincide. In this limit the phase-space 
overlap between the paired states is maximal.  
Consider the effect of a finite $E_A$ when $P=0=\delta\epsilon$. It 
increases/decreases the energies of the isospin up/down particles; 
the  shift in the Fermi-levels 
of the isopin up/down particles samples different momentum
regions of the phase-space in the kernel of the gap equation. This
blocking effect or  phase-space decoherence in the pair states in turn
reduces the magnitude of the gap. In particular this allows for 
pairs with small $E_A$. Switching on a finite
$P$ and/or $\delta\epsilon >0$ acts to restore partially the phase space 
coherence of the isospin up/down states when $\delta\mu \ge{\rm max}\{
Pp/2m,~  \mu\delta\epsilon\}$ [see Eq. (\ref{EA})]. 
Therefore, the
magnitude of the gap increases in this case, i.e. the phase-space
overlap of the Fermi surfaces is (partially) restored. Increasing 
$P$ and/or $\delta\epsilon >0$ further acts in a manner similar 
to pure energy shift $\delta\mu$  - the decoherence increases and the 
pairing is eventually destroyed. 
This picture is seen in Fig. 1 in the region $\delta\epsilon >
0$. Small perturbations in $P$ and $\delta\epsilon$ from the
asymmetric BCS state ($ P=0=\delta\epsilon$ but $\delta\mu \neq 0$)
increase the pairing gap. For large perturbation in either $P$ {\it or} 
$\delta\epsilon > 0$ the gap  vanishes as the decoherence
increases. The combined effect 
of the finite momentum and the quadrupole deformations is seen
in the minimum plateau in the region of small $P$ and  $\delta\epsilon$, 
followed by a maximum in the gap with increasing $P$ and/or  $\delta\epsilon$,
and a rapid falloff beyond the maximum. The behavior of the gap in the
limits of pure LOFF and DFS phases is the same as the one found in the 
previous work \cite{NUCL_LOFF,MS}. 

In the case of  $\delta\epsilon< 0$  
the $\mu\delta\epsilon$ term in Eq. (\ref{EA}) 
combines with the chemical potential shift $\delta\mu$. To
satisfy a given particle number asymmetry there are now two parameters
(while there was only one, $\delta\mu$, for positive $\delta\epsilon$). 
As a result with increasing deformation and hence 
$\mu\delta\epsilon$ (the changes in $\mu$  are insignificant) the
density asymmetry can be maintained at a cost of smaller
$\delta\mu$ which tends to zero and eventually changes the sign
($\mu_p > \mu_n)$ at $\delta\epsilon \sim -1.$
This marks the  ``turnover'' in the gap equation which eventually
vanishes for $\delta\epsilon > -1.$ We shall see that the regions of
the negative values of the deformation parameter are physically not
relevant, since these states have larger kinetic energies. As in the 
case of positive $\delta\epsilon$ the effect of finite momentum is to
suppress the pairing for a large deformation; the combined phase
increases the pairing only for small perturbations from the asymmetric
BCS state in the vicinity of the ``minimum plateau'' ($\delta\epsilon\to
0, $  $P\to 0$). The symmetric part of the quasiparticle spectrum (\ref{ES}) 
controls the location of the maximum (corresponding to $E_S = 0$)  and
the width of the kernel in Eq. (\ref{GAP4}) in the momentum space 
through its contribution to the denominator of the gap equation. A
positive contribution to the average chemical potential effectively
decreases the 
density of states  (which can be easily seen
analytically in the zero temperature and weak coupling limits), and
therefore, the gap function.
\begin{figure}[htb] 
\begin{center}
\includegraphics[angle=-90,width=\linewidth]{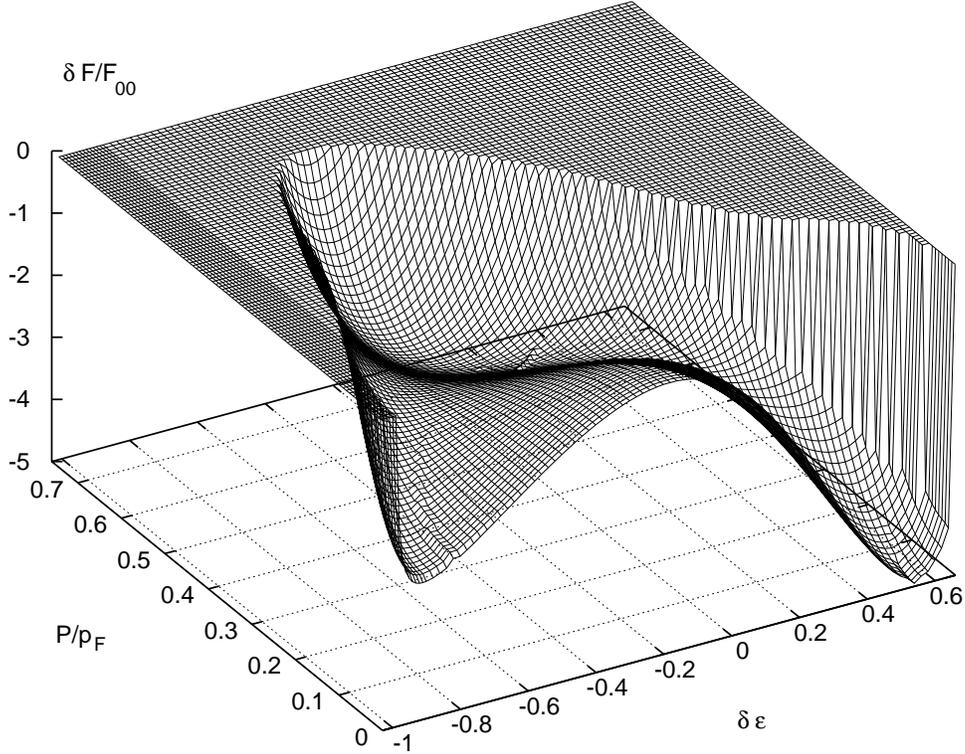}
\end{center}
\caption{The free energy difference between the superconducting and 
normal states $\delta {\cal F}$  as a function of the Cooper pair 
momentum $P$ in units of Fermi momentum and the relative 
deformation parameter $\delta\epsilon$. The free energy is normalized
to its value $\delta{\cal F}_{00}$ in the asymmetric BCS state.
The parameters are the same as in Fig. 1.
}
\end{figure}
\noindent
 Inversely, a negative 
contribution to the average chemical increases the density of states. 
The changes in the $E_S$ due to the deformation do not affect
qualitatively the
results above. In summary of  Fig. 1, the combination of the LOFF 
and DFS phases promotes the pairing only in the region of small 
$\delta\epsilon$ and $P$ (weak perturbation from the BCS state). For large 
quadrupole deformations the finite momentum is not favorable. 
Viewed from the LOFF phase, for large values of $P$ the quadrupole
deformations are disfavored.

Figure 2 displays the difference between the free energies of the
superconducting and normal states $\delta{\cal F}$ normalized to its
value in the asymmetric BCS state $\delta{\cal F}_{00}=\delta{\cal
F}(P=0, \delta\epsilon = 0)$. The thermal
contribution due to the finite-temperature entropy is numerically 
irrelevant for the net budget of free energies of  both states. 
Whether the superconducting state is thermodynamically favored depends
on the relative magnitude of the potential energy of the pair
interactions [second term in Eq. (\ref{U})] and the difference in the
kinetic energies of the normal and superconducting states. Since the
energy of the pair interactions scales as the square of the pairing
gap, the shape of the $\delta{\cal F}$ surface closely resembles 
that of the pairing gap in Fig. 1. There are, however,
significant quantitative differences due to the contribution from the
kinetic energy of the quasiparticles. The asymmetric BCS state is the
stable ground state of the system (${\cal F} < 0$), however its
perturbations for finite $\delta\epsilon$ and $P$ are unstable towards
evolution to lower energy states. For the  pure LOFF phase
($\delta\epsilon = 0$)  
the ground state corresponds to finite momentum $P\sim
0.5$ (in units of $p_F$). 
For the pure DFS phase ($P = 0$) there are two
minima corresponding to $\delta\epsilon \simeq -0.8$ and 
$\delta\epsilon\simeq 0.55$, i.e., prolate and oblate deformations of
neutron Fermi spheres, respectively.  In  general the
position of the minimum of $\delta{\cal F}$ in the
$\delta\epsilon$-$P$ plane (passing through the minima of the
limiting cases) prefers either large deformations or large finite
momenta. The absolute minimum energy state corresponds to
$\delta\epsilon\simeq 0.55$ {\it and } $P = 0$; that is, while the
LOFF phase is a local minimum state, it is generally unstable towards
evolution to a pure DFS phase with oblate/prolate deformations of
neutron/proton Fermi spheres. The effect of the kinetic energy
contribution is to suppress the large pairing contribution for
negative values of the deformation parameter. The position of the 
true minimum of the free energy coincides with the maximum in the
pairing gap in the positive $\delta\epsilon$ regime.

\subsection{LOFF phase at finite-temperature}

When the system is in an isospin asymmetric state the effect of the 
temperature on the pairing gap and related characteristics of the
superconducting state is twofold: 
first, at high temperatures 
the pairing is suppressed due to the thermal excitation of the
quasiparticle states, in analogy to the classical BCS superconductors;
second, at low temperatures the 
\begin{figure}[b] 
\begin{center}
\includegraphics[angle=-90,width=\linewidth]{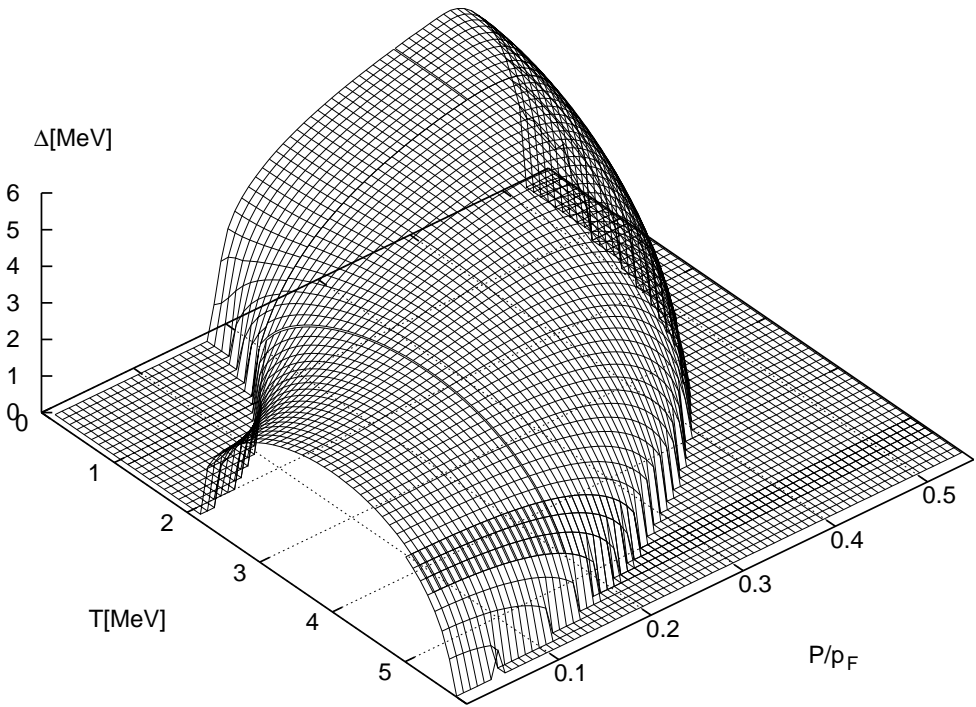}
\end{center}
\caption{The LOFF pairing gap as a function of the Cooper pair 
momentum and the temperature. The parameter values are the same as in
Fig. 1.
}
\label{MSfig:fig4}
\end{figure}
\noindent  
pairing  is enhanced by the temperature,
since the thermal excitation of the quasiparticle states at different
Fermi surfaces reduces the blocking effect and 
increases the phase-space overlap. Figure 3 shows the
pairing gap in the LOFF state as a function of the temperature and
finite momentum.
For $P=0$ the asymmetric BCS state exists between
two critical temperatures corresponding to the thermal stimulation and
suppression of the pairing. For constant
asymmetry the temperature reduces the antisymmetric part 
of the quasiparticle spectrum [$E_A$ in Eq. (\ref{GAP4})],
and hence the decoherence between the quasiparticle states 
on different Fermi surfaces. This is responsible for the gap 
{\it reentrance} phenomenon  at lower critical temperature with 
increasing temperature. The reduction of  the
contribution from the symmetric part $E_S$ of the spectrum with
increasing temperature is responsible for the quenching of the
superconducting state at the upper critical temperature. At $T=0$,
but for finite $P$,  analogous reentrace of the gap function is seen in
Fig. 3. The existence of the lower and upper critical momenta can
again be understood from the contributions of the finite momentum to the
quasiparticle spectrum.
\begin{figure}[htb] 
\begin{center}
\includegraphics[angle=-90,width=\linewidth]{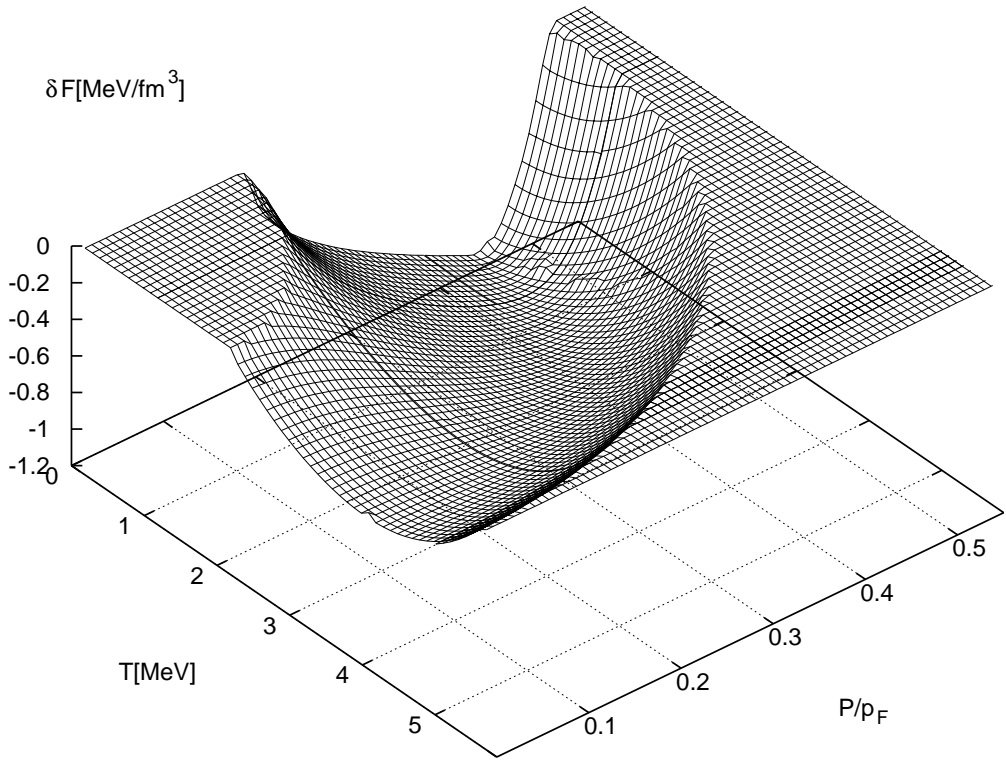}
\end{center}
\caption{The free energy difference between the superconducting and 
normal states $\delta {\cal F}$  as a function of the Cooper  pair 
momentum $P$ in units of Fermi momentum and temperature.
The parameters are the same as in Fig. 1.
}
\label{MSfig:fig5}
\end{figure} 
In the general case  $P\neq 0$ and $T\neq 0$  
the reentrace phenomenon extends 
in the $P-T$ plane: the pairing region of
the phase diagram is bounded by the curves for the lower and 
upper critical values $P_{c1}(T)$, $P_{c2}(T)$, or equivalently 
$T_{c1}(P)$, $T_{c2}(P)$. The duality between the effect of finite $T$ 
on the pairing gap at fixed $P$ to that of  finite $P$  at fixed
$T$ is evident. 
In Fig. 4 we show the difference between the free energies of the
superconducting and normal states $\delta{\cal F}$ for the LOFF phase. 
The true ground state here corresponds to the zero-temperature limit 
for pair momentum $P/p_F\sim 0.4$. The effect of the temperature on the
LOFF phase depends on the momentum of the pair: for low momenta there
is a minimum as a function of the temperature (as is also the case
for the asymmetric BCS state). For large fixed momenta the
temperature reduces the pairing gap. The same applies for the
perturbations from the $P=0$ state at fixed temperature: for low
temperatures the system prefers finite momenta, while for large
temperatures the finite momentum only reduces the energy. The overall
dependence of the free energy  on the temperature and the finite
momentum reflects the dependence of the pairing gap and the
potential energy of the pairing interaction on these quantities.

\subsection{DFS phase at finite-temperature}

The features seen in the phase diagram of the DFS phase at finite
temperatures are analogous
to those already discussed for the LOFF phase. Here we reiterate this 
discussion with the emphasis on the qualitative differences of these
phases. The pairing gap as a function of the temperature and the
relative deformation ($\delta\epsilon$)  is shown in Fig. 5. 
\begin{figure}[htb] 
\begin{center}
\includegraphics[angle=-90,width=\linewidth]{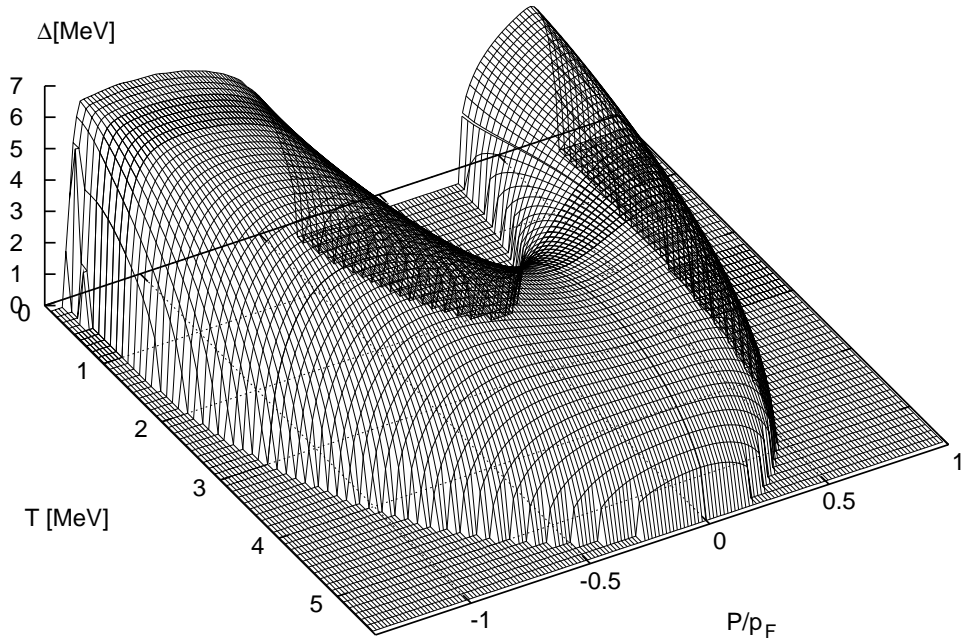}
\end{center}
\caption{The DFS pairing gap as a function of the relative deformation
 and the temperature. The parameter values are the same as in
Fig. 1.
}
\label{MSfig:fig7}
\end{figure}
\noindent 
The reentrace effect for $T = 0$ is seen for both signs of
$\delta\epsilon$; the lower critical deformation corresponds to the
onset of the pairing due to the restoration of the phase-space
coherence via the deformation of the spherical Fermi surface into
quasiellipsoidal form. The upper critical $\delta\epsilon$
corresponds to the loss of coherence due to the large shift in the
Fermi surfaces of the paired particles. The $T-\delta\epsilon$ duality
is seen in the case where deformation vanishes (asymmetric BCS state).
The lower critical temperature corresponds to the thermal stimulation
of the pair correlations due to the smearing of the Fermi surfaces,
while the upper critical temperature corresponds to the ordinary
thermal suppression of the pairing in the BCS-normal state transition.
The effect of $\delta\epsilon$ at finite-temperatures can be
understood in terms of its contribution to the antisymmetric  part of
the quasiparticle spectrum, in complete analogy to the LOFF
phase. The reentrace effect extends in the plane $T-\delta\epsilon$,
and the pairing region is bounded by the critical values 
$\delta\epsilon_1(T)$ and $\delta\epsilon_2(T)$, or equivalently 
$T_{c1}(\delta\epsilon)$ and $T_{c2}(\delta\epsilon)$. Unlike the LOFF
phase these curves do not terminate at $\delta\epsilon = 0$, rather they
show a continuous crossover from the oblate to the prolate
deformation of the Fermi surface of the paired particles. For either
sign of $\delta\epsilon$ the temperature thermally stimulates the
superconducting phase for small perturbations from the asymmetric BCS
state corresponding to the $\delta\epsilon = 0$ cut; 
for large deformations the pairing is largest in the 
zero-temperature limit. 

\begin{figure}[htb] 
\begin{center}
\includegraphics[angle=-90,width=\linewidth]{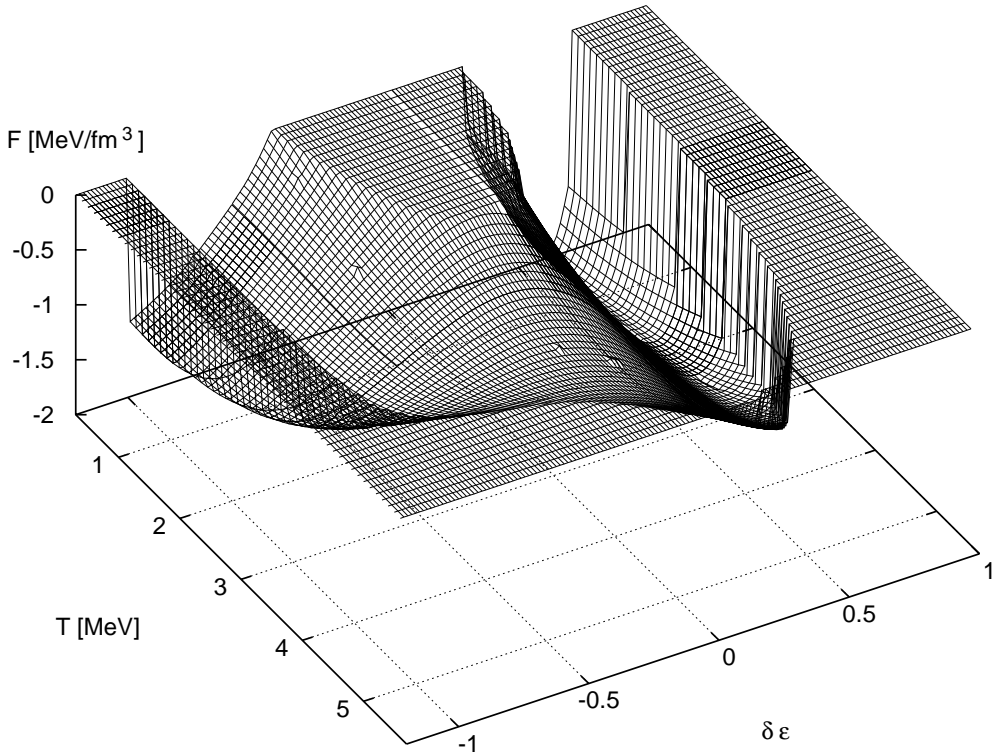}
\end{center}
\caption{The free energy difference between the superconducting and 
normal states $\delta {\cal F}$  as a function of the relative 
deformation and temperature. The parameters are the same as in Fig. 1.
}
\label{MSfig:fig9}
\end{figure} 
The effect of finite $\delta\epsilon$ at fixed
temperature is a suppression of the pairing in the high-temperature
limit, while at lower temperatures there is a pronounced maximum in
the pairing gap as a function of $\delta\epsilon$.
Finally, the difference between the free energies of the
superconducting and normal states $\delta{\cal F}$ for the DFS phase
is shown in Fig. 6. There are two minima corresponding to the
oblate or prolate deformations of the Fermi surfaces of 
neutrons/protons. The shape of $\delta{\cal F}$ is controlled by the
interplay between the pairing potential energy and the kinetic energy;
the contribution from the entropy difference is numerically
negligible. The minima correspond to large deformations of the order 
0.5 which optimize the gain in the potential energy  over the 
loss in the kinetic energy due to the deformation; the true minimum
occurs for positive $\delta\epsilon$ at zero-temperature. While
temperature stimulated pairing correlations are seen for small
deviations from the BCS state, this effect is not essential, since the
minimum of the $\delta{\cal F}$ surface occurs at large 
$\delta\epsilon$. The DFS phase is most stable in the 
zero-temperature limit.

\section{Summary and outlook}
The fermionic condensation in nuclear matter under isospin asymmetry 
leads to superconducting states which spontaneously break the space
symmetries. We studied two realizations of such condensates - the LOFF
phase, where the Cooper pairs carry finite momentum, and the DFS phase
where the Fermi surfaces of the paired nucleons are deformed, to the
lowest order  into
ellipsoidal form. A unifying feature of these phases is the
deformation of the spectrum of the quasiparticle excitations at the 
leading (LOFF) and next-to-leading (DFS) order in the expansion in
Legendre polynomials  with respect to the angle between the particle
momenta and the  axis of symmetry breaking.
When these deformations are treated simultaneously, we find that the
pure DFS phase has lower energy than the LOFF phase. The combination
of these phases still lower the energy of the normal system and can
exists in a metastable state. In particular, the pure LOFF state could be 
realized in the case where 
the unpaired state features some intrinsic spatial asymmetries
(e.g., due to the lattice in the metallic superconductors).
Of course, the axis of the symmetry breaking for the LOFF and DFS
phases need not  be the same and, more generally, the patterns of symmetry 
breaking could be more complicated than those studied here (e.g., 
higher order multipole deformations of the Fermi surfaces). 

The phase diagrams of the LOFF and DFS phases in the
temperature/deformation plane show a number of unusual features: 
(i) two critical temperatures for the onset of the superconducting
state (the reentrace phenomenon). The upper critical temperature is
the analog of the BCS temperature for the superconducting-normal phase 
transition. The lower critical temperature is specific to asymmetric
systems and results from the thermally stimulated phase-space overlap
between the pairing states. (ii) There exists duality between the temperature
effects and the finite-momentum or deformation effects. For example,
at low temperatures the LOFF phase is characterized by two critical
momenta of the Cooper pairs at which the superconducting phase sets in.
(iii) 
The free energy of the combined LOFF and DFS
phases shows local minima corresponding to negative and positive 
relative deformations at zero momentum of the pairs. The lowest energy
corresponds to a phase with oblate neutron and prolate proton Fermi
surfaces ($\delta\epsilon$ is positive).

The realization of the $^3S_1-^3D_1$ pairing in  bulk nuclear
matter depends on a number of factors: (i) the ratio of the pairing
gap to the average chemical potential, which controls the
density-temperature-dependent critical asymmetries at which the
pairing vanishes; this ratio depends sensitively on the
quasiparticle renormalization effects which need  further study in
this context. (ii) A second factor involves the asymmetries implied by the
$\beta$ equilibrium of the matter with leptons in the stellar matter;  
while the asymmetries derived for the normal state may serve as
order of magnitude estimates, the LOFF and DFS phases under 
$\beta$ equilibrium might require asymmetries largely different from
those deduced for the unpaired system. (iii) The matter could prefer a
state where the superconducting and normal phases coexist in a mixed
phase, with superconducting ``islands'' immersed in the unpaired matter;
the analogy to the type-I superconductors in the magnetic fields is a
useful counterpart of this type of ordering. (iv) The extreme 
low-density asymmetric nuclear matter can support pairing in the 
$^3S_1-^3D_1$ channel at much larger asymmetries than the matter, 
say, at the nuclear saturation density; however, in this strong coupling
regime one deals with a Bose-Einstein condensate of deuterons
coexisting with a low-density neutron gas\cite{NUCL_BCS3} and one
needs an understanding of the  crossover from the LOFF and DFS phases 
to their Bose-Einstein condensed counterparts.

We now briefly comment on possible realization of the phases with
broken space symmetries in other nuclear systems. The neutron-proton
pairing in finite nuclei\cite{RS,AG_REVIEW,WFS,AG99,Engel96,Engel97,Doba96,Satula97,Civ97,Civ97_bis,Roepke00} 
is an obvious candidate. The
odd-number projected BCS equations contain the decoherence
effects in the gap equation introduced by the extra odd particle. The
resulting blocking effect is quite analogous to the 
phase decoherence due to the imposed neutron-proton asymmetry in the
bulk nuclear matter.
The problem of realization of the LOFF phase is facilitated by the
fact that the pairing instability occurs for total momenta  
$P\sim p_F\gg 1/R$, where $R$ is the radius of the
nucleus. Therefore, for already medium size nuclei the influence of
the surface on the condensation energy should not be large and the
solutions found for the interior of the nucleus should go over to those
corresponding to infinite systems. The translational symmetry is
not broken by the DFS phase, therefore the finite size effects (while
being generally of the same order of magnitude as for the LOFF phase)
would affect the structure of the condensate to the same extent as
those for the case of the pairing with undeformed Fermi surfaces. 

Quite generally, the phases with broken space symmetries arise in the
physical situations where the time-reversal symmetry between the
paired fermion states is broken. This is the case, for example, in
finite nuclei under rotation. The $^1S_0$ pairing among
neutrons and/or protons in such nuclei can support the LOFF or
DFS like states.
Similarly, the magnetic field breaks the time-reversal
invariance due to the Pauli paramagnetism of neutrons and protons.
The LOFF and DFS phases can arise for the case of the 
 $^1S_0$ pairing in strong magnetic fields, for example, in neutron
star crusts. These phases will emerge when the interaction energy of
a neutron spin with the magnetic field $\mu_BB$, where $\mu_B$ is the 
anomalous magnetic moment of a neutron, is of the order
of the pairing gap in the zero field limit. This condition is
satisfied for the field strength of the order $10^{16}$-$10^{18}$ G (depending
on the value of the pairing gap) which could occur in the highly
magnetized subpopulation of neutron stars (magnetars).

As is well know, the breaking of the continuous symmetries in
the infinite systems leads to collective excitations with vanishing
minimal frequency (Goldstone's theorem). The LOFF phase breaks the
translational and rotational symmetries; the existence of the
preferred direction in the DFS phase breaks the rotational symmetry of the 
system. The breaking of continuous symmetries in the ground of these phases,
therefore, implies new collective bosonic 
modes in asymmetric nuclear matter. 
In finite systems the Goldstone theorem implies collective
modes which tend to zero as a certain power of the radius of the
system. The observation of the Goldstone modes associated with the
LOFF and/or DFS phases could provide  information on the true ground
state structure of the condensate under laboratory conditions, for
example,  in finite nuclei.

\section*{Acknowledgments}
We would like to acknowledge the financial support by the 
Sonderforschungsbereich 382 of DFG.

\end{document}